\documentclass[a4paper,american,aps,amsfonts,amsmath,prd,tightenlines, nofootinbib]{revtex4}
\usepackage[T1]{fontenc}
\usepackage[latin1]{inputenc}
\usepackage{amssymb}
\makeatletter
\usepackage{epsf}

\def\be{\begin{equation}}
\def\ee{\end{equation}}
\def\beq{\begin{equation}}
\def\eeq{\end{equation}}
\def\bea{\begin{eqnarray}}
\def\eea{\end{eqnarray}}
\def\bml{\begin{subequations}}
\def\blea{\bml\begin{eqnarray}}
\def\elea{\end{eqnarray}\end{subequations}}

\usepackage{babel}
\makeatother
\begin{document}

\title{Cosmological perturbations from vector inflation}

\author{Alexey Golovnev}

\author{Vitaly Vanchurin}

\affiliation{Arnold-Sommerfeld-Center for Theoretical Physics, Department f\"{u}r
Physik, Ludwig-Maximilians-Universit\"{a}t M\"{u}nchen, Theresienstr.
37, D-80333, Munich, Germany}

\begin{abstract}
We analyze the behavior of linear perturbations in vector inflation.
In contrast to the scalar field inflation, the linearized theory with
vector fields contains couplings between scalar, vector and tensor
modes. The perturbations decouple only in the ultraviolet limit, which
allows us to carry out the canonical quantization. Superhorizon perturbations
can be approximately analyzed due to suppressed mixing between different
modes in the small fields models. We find that the vector perturbations
of the metric decay exponentially, but the scalar and tensor modes
could remain weakly coupled throughout the evolution. As a result,
the vector inflation can produce significant correlations of the scalar
and tensor modes in the CMB. For the realistic models the effect is
rather small, but not negligible.
\end{abstract}
\maketitle

\section{Introduction}

Theory of inflation relates the laws of physics at very small distances
to observations at extremely large scales. An essential ingredient
of the standard inflationary paradigm is a scalar field \cite{Linde}
which looks like a disturbing restriction, especially because scalar
fields are known not to be abundant in nature. In an attempt to cure
the problem we proposed a new model of inflation where the quasi de
Sitter expansion is driven by vector fields \cite{GMV}. It seems
as the most natural step beyond the scalars, although the possibility
of inflation with other fields were also analyzed in the literature.
For example, the spinors \cite{Wata}, the dark spinors \cite{boeh},
or even the spin fluid in the Einstein-Cartan theory with torsion
\cite{brr} can be employed to drive quasi de Sitter expansion. The
most interesting generalization of the vector inflation \cite{GMV}
to higher forms was implemented in Ref. \cite{Germani}.

To construct isotropic solutions one can consider purely time-like
vector fields \cite{harko,Mota,KohHu,SKoh} which do not behave like
genuine vectors but rather resemble some modified scalars. More realistically,
an approximate isotropy can be achieved with a triad of mutually orthogonal
vector fields \cite{Armendariz} or with a large number of randomly
oriented fields \cite{GMV}. The well-known problem of the slow-roll
\cite{Ford} was resolved \cite{GMV} by non-minimal coupling of vector
fields to gravity. In general, the model can be described by the following
action\begin{align}
S & =\int\sqrt{-g}\left[-\frac{R}{2}\left(1+\sum_{n=1}^{N}\frac{1}{6}I_{(n)}\right)-\frac{1}{4}\sum_{n=1}^{N}F_{\mu\nu}^{(n)}F_{(n)}^{\mu\nu}-\sum_{n=1}^{N}V\left(I_{(n)}\right)\right]dx^{4}.\label{eq:action1}\end{align}
where $I_{(n)}\equiv{-A}_{\mu}^{(n)}A_{(n)}^{\mu}$ and $F_{\mu\nu}^{(n)}\equiv\nabla_{\mu}A_{\nu}^{(n)}-\nabla_{\nu}A_{\mu}^{(n)}$.
In a spatially flat Friedmann universe the evolution of the homogeneous
background fields in conformal coordinates is described by\begin{align}
A_{0} & =0\nonumber \\
B_{i}^{\prime\prime}+2{\cal H}B_{i}^{\prime} & +2V_{,I}a^{2}B_{i}=0\label{eq:field}\end{align}
where $I=B_{i}B_{i}\equiv B^{2}$,$B_{i}\equiv\frac{A_{i}}{a},$ and
${\cal H}\equiv\frac{a'}{a}$, and the Einstein equations can be written
as \begin{equation}
{3{\cal H}}^{2}=8\pi N\left(Va^{2}+\frac{B^{\prime2}}{2}\right),\label{eq:einstein}\end{equation}
\begin{equation}
2{\cal H}^{\prime}+{\cal H}^{2}=8\pi N\left(Va^{2}-\frac{B^{\prime2}}{2}\right).\label{eq:einstein2}\end{equation}

From the point of view of the background evolution, the proposed model
is very similar to the scalar field inflation. However, in contrast
to the standard inflation, the expansion in vector inflation must
no longer be isotropic, which could lead to very distinct observational
predictions. The current observational bounds on isotropy of inflation
are very weak and one can easily allow $\sim10\%$ of anisotropy.
For example, the vector inflation with $N\sim100$ randomly oriented
vector fields can give rise to anisotropy of the order of $\frac{1}{\sqrt{N}}\sim10\%$
at the end of inflation. The anisotropy is washed out shortly after
the end of inflation when the energy-momentum tensor becomes isotropic,
but the signatures of anisotropic inflationary stage might still be
observable.

A somewhat closer look at the proposed scenario had shown that models
of large fields vector inflation (e.g. $V=-\frac{m^{2}A_{\mu}A^{\mu}}{2}$)
are generically unstable due to tachyonic behavior of gravitational
waves \cite{GMV-grav}. Nevertheless, the models of the small fields
inflation (i.e. Coleman-Weinberg potential) are stable under tensor
perturbations of the metric. The main objective of the current paper
is to investigate the behavior of a general type of linear perturbations
in vector inflation.%
\footnote{Some problems related to the linear perturbations in vector inflation
have been already studied in Refs. \cite{Chiba,Lyth}. See also \cite{Shank,Koba}
for linear perturbation analyses in other non-scalar inflationary
models.%
} However, the proposed analysis can be applied to the model of inflation
with vector impurity \cite{Soda} and to the models with vector curvaton
\cite{dimop}.

The article is organized as follows. In the next section we derive
the linearized equations of motion. In the third section we concentrate
on the ultraviolet limit and quantize the perturbations of vector
fields. The evolution of the long wavelength modes is analyzed in
the forth section for the small coupling limit. In the final section
we summarize the main results and discuss observational consequences
of the correlations between scalar and tensor modes.

\section{Linearized equations}

To study the evolution of cosmological perturbations we consider small
variations of the metric around a spatially flat Friedmann background.
In conformal coordinates the line element is given by \[
ds^{2}={a\left(\eta\right)}^{2}\left(\left(1+2\phi\right)d\eta^{2}+2{\cal V}_{i}d\eta dx^{i}-\left(\left(1-2\psi\right)\delta_{ik}-h_{ik}\right)dx_{i}dx_{k}\right)\]
where ${\cal V}_{,i}^{i}=0$, $h_{i}^{i}=0,$ $h_{j,i}^{i}=0$. General
perturbations of the vector fields are described by \begin{equation}
\delta A_{\alpha}=\left(\delta A_{0},\chi_{,i}+{\delta A}_{i}^{T}\right)\label{eq:longitudinal}\end{equation}
where ${\delta A}_{,i}^{Ti}=0$. Overall there are $2+2N$ scalar
($\phi$, $\psi$, $\delta A_{0}$ and $\chi$), $2+2N$ vector ($V_{i}$
and ${\delta A}_{i}^{T}$) and $2$ tensor ($h_{ij}$) variables. 

The perturbed Ricci scalar and Einstein tensor are given by \begin{equation}
a^{2}\delta R=6\psi''-4\bigtriangleup\psi+2\bigtriangleup\phi+6{\cal H}\left(\phi'+3\psi'\right)+24\phi{\cal H}^{2},\label{eq:Ricci}\end{equation}

\begin{align}
a^{2}\delta G_{0}^{0} & =2\bigtriangleup\psi-6{\cal H}\psi'-6{\cal H}^{2}\phi\nonumber \\
a^{2}\delta G_{k}^{0} & =2\left(\psi'+{\cal H}\phi\right){}_{,k}+\frac{1}{2}\bigtriangleup{\cal V}_{k}\label{eq:Gtensor}\\
a^{2}\delta G_{k}^{i} & =\left(-2\psi''+\bigtriangleup\left(\psi-\phi\right)-2{\cal H}\left(\phi'+2\psi'\right)-6\phi{\cal H}^{2}\right)\delta_{k}^{i}+\left(\phi-\psi\right)_{,ik}\nonumber \\
 & -\frac{1}{2}{{\cal V}'}_{\{ i,k\}}-{\cal H}{\cal V}_{\{ i,k\}}+\frac{1}{2}{h''}_{ik}+{\cal H}{h'}_{ik}-\frac{1}{2}\bigtriangleup h_{ik}.\nonumber \end{align}
Variation of the energy-momentum tensor is a straight-forward but
rather bulky exercise. For an isotropic background we have $\sum_{n=1}^{N}A_{i}^{(n)}A_{k}^{(n)}\propto\delta_{ik}$
and the equations could be somewhat simplified since $\left\langle A_{i}A_{k}\right\rangle =\left\langle A_{i}^{2}\right\rangle \delta_{k}^{i}=\frac{1}{3}\langle B^{2}\rangle\delta_{k}^{i}$.
This assumption also implies $\sum_{n=1}^{N}h_{ik}A_{i}^{(n)}A_{k}^{(n)}=0$,
since $h_{ik}$ is traceless. Linear terms in the Einstein equations
read as (see Appendix A):

\begin{eqnarray}
\frac{a^{4}}{N}\delta T_{0}^{0} & = & \left\langle 2\left(\frac{1}{2}{A'}_{i}^{2}+\frac{1}{2}{\cal H}^{2}A_{i}^{2}-{\cal H}A_{i}{A'}_{i}\right)\left(\psi-\phi\right)-{\cal H}\psi'A_{i}^{2}+\psi'A_{i}{A'}_{i}+{2a}^{2}V_{,I}{\psi A}_{i}^{2}\right.\label{eq:T00}\\
 &  & -{A'}_{i}{\delta A}_{0,i}+{\cal H}^{2}A_{i}\delta A_{i}+{2a}^{2}V_{,I}\left(A_{i}\delta A_{i}\right)-\left.{\cal H}\left(A_{i}\delta A_{i}\right)'+A_{i}'\delta A_{i}'+\frac{1}{3}A_{i}\bigtriangleup\delta A_{i}\right\rangle \nonumber \end{eqnarray}

\begin{eqnarray}
\frac{a^{4}}{N}\delta T_{k}^{0} & = & \left\langle A'_{i}{\delta A}_{[i,k]}^{T}\right.-\left(\frac{a''}{a}-2a^{2}V_{,I}\right)A_{k}\left({\delta A}_{0}+A_{i}{\cal V}_{i}\right)-\frac{1}{12}A_{i}^{2}\bigtriangleup{\cal V}_{k}\nonumber \\
 &  & +\left.\left(-{\cal H}A_{i}^{2}\psi+\frac{1}{3}A_{i}{A'}_{i}\left(2\psi-\phi\right)+\frac{1}{3}\left(A_{i}\delta A_{i}\right)^{\prime}-{\cal H}A_{i}\delta A_{i}\right)_{,k}\right\rangle \label{eq:T0k}\end{eqnarray}

\begin{eqnarray}
\frac{a^{4}}{N}\delta T_{k}^{i} & = & \left\langle \left[2\left(-\frac{5}{6}{A'}_{j}^{2}-\frac{1}{2}{\cal H}^{2}A_{j}^{2}-\frac{1}{3}A_{j}{A''}_{j}+{\cal H}A_{j}{A'}_{j}\right)\left(\psi-\phi\right)\right.+\frac{1}{6}A_{j}^{2}\bigtriangleup\left(\phi+\psi\right)+\frac{1}{3}A_{j}{A'}_{j}\left(\phi'-2\psi'\right)\right.\nonumber \\
 &  & +{\cal H}A_{j}^{2}\psi'+2\left(\frac{a''}{a}-2a^{2}V_{,I}\right)A_{i}^{2}\psi+2{A'}_{i}^{2}\left(\psi-\phi\right)+{A'}_{j}{\delta A}_{0,j}+\left({\cal H}A-A_{j}'\right)\delta A_{j}'-\frac{1}{3}\left(A_{j}\delta A_{j}\right)''\label{eq:Tik}\\
 &  & +{\cal H}\left(A'{}_{j}-{\cal {\cal H}}A_{j}\right)\delta A_{j}+\frac{1}{3}A_{j}\bigtriangleup\delta A_{j}+A_{i}^{2}\left(\frac{1}{3}\bigtriangleup\left(2\psi-\phi\right)-{\cal H}\left(3\psi'+\phi'\right)-2\frac{a''}{a}\phi-\psi''\right)\nonumber \\
 &  & +\left.2a^{2}V_{,I}\left({\psi A}_{l}^{2}+A_{l}\delta A_{l}\right)\right]\delta_{k}^{i}-\frac{1}{6}A_{j}^{2}\left(\psi+\phi\right)_{,ik}-4A_{i}A_{k}V_{,II}\left({\psi A}_{l}^{2}+A_{l}\left(\delta A_{l}+\frac{1}{2}h_{lj}A_{j}\right)\right)\nonumber \\
 &  & +\left(\frac{a''}{a}-2a^{2}V_{,I}\right)\left(A_{\{ i}{\delta A_{k}}_{\}}+h_{ij}A_{k}A_{j}\right)-\frac{1}{3}A_{j}{\delta A_{j}}_{,ik}+{A'}_{\{ i}\delta A_{k\}}'-A'_{\{ i}{\delta A}_{0,k\}}\nonumber \\
 &  & +\left.\frac{1}{6}A_{j}{A'}_{j}\left({\cal V}_{\{ i,k\}}-{h'}_{ik}\right)+\frac{1}{12}A_{j}^{2}\left({\cal V}_{\{ i,k\}}-{h'}_{ik}\right)'+\frac{1}{12}A_{j}^{2}\bigtriangleup h_{ik}+{A'}_{k}{A'}_{j}h_{ij}\right\rangle \nonumber \end{eqnarray}
where the summation over $j$ is implied. $V$, $V_{,I}$ and $V_{,II}$
are evaluated at the background values of $I=B^{2}$ and the terms
with $V_{,II}$ can be further simplified by careful averaging. 

The perturbed equations of motion for each of the vector fields are

\begin{equation}
\left(\phi+\psi\right)_{,i}{A'}_{i}+\bigtriangleup\left(\delta A_{0}-\chi'\right)+\left(\frac{a''}{a}-2a^{2}V_{,I}\right)\left(\delta A_{0}+{\cal V}_{i}A_{i}\right)=0\label{eq:EOM01}\end{equation}

\begin{eqnarray}
-2\phi{A''}_{i}-\left(\phi'+\psi'\right){A'}_{i}-{\delta A_{0}}_{,i}'-\bigtriangleup{\delta A}_{i}^{T}-{A'}_{k}{\cal V}_{i,k}+{h'}_{ik}{A'}_{k}+\delta A_{i}''\label{eq:EOM02}\\
+\left(2a^{2}V_{,I}-\frac{a''}{a}\right)\delta A_{i}+2\frac{d^{2}V}{dI^{2}}A_{i}\left({2\psi A}_{l}^{2}+2A_{l}\delta A_{l}+A_{l}A_{j}h_{lj}\right)\nonumber \\
+\left(\psi''+\frac{1}{3}\bigtriangleup\left(\phi-2\psi\right)+{\cal H}\left(\phi'+3\psi'\right)+2\phi\frac{a''}{a}\right)A_{i} & = & 0.\nonumber \end{eqnarray}
Equations (\ref{eq:T00}), (\ref{eq:T0k}), (\ref{eq:Tik}), (\ref{eq:EOM01})
and (\ref{eq:EOM02}) form a very complicated, but closed system.
In what follows we analyze the perturbations in the ultraviolet (Section
III) and infrared (Section IV) limits. In the Appendix B, for the
purpose of completeness, we consider purely adiabatic perturbations
which are not consistent with equations of motion.

\section{quantum perturbations }

In the short wavelength limit the terms containing less than two derivatives
vanish and Eqs. (\ref{eq:EOM01}) and (\ref{eq:EOM02}) imply \begin{equation}
\delta A_{0}=\chi^{\prime}\label{eq:D-E}\end{equation}

\[
\chi_{,i}^{\prime\prime}-{\delta A}_{0,i}^{\prime}+{{\delta A}_{i}^{T}}^{\prime\prime}-\bigtriangleup{\delta A}_{i}^{T}+\left(\psi^{\prime\prime}+\frac{1}{3}\bigtriangleup\left(\phi-2\psi\right)\right)A_{i}=0,\]
where the later equation can be decomposed into vector and scalar
parts: \[
{{\delta A}_{i}^{T}}^{\prime\prime}-\bigtriangleup{\delta A}_{i}^{T}=0\]
\begin{equation}
\psi^{\prime\prime}+\frac{1}{3}\bigtriangleup\left(\phi-2\psi\right)=0.\label{eq:psi-phi}\end{equation}

The $00$ component of the Einstein equation

\[
\frac{2}{a^{2}}\bigtriangleup\psi\approx\sum_{n=1}^{N}\frac{1}{3a^{4}}A_{i}^{(n)}\bigtriangleup\delta A_{i}^{(n)}\]
yields \begin{equation}
\psi=\frac{1}{6a^{2}}\sum_{n}A_{i}^{(n)}\delta A_{i}^{(n)}=\frac{1}{6}\sum_{n}B_{i}^{(n)}\delta B_{i}^{(n)}\label{eq:psi-BdB}\end{equation}
and the $ik$ component gives the usual wave equation for gravitational
waves together with an unusual condition \begin{equation}
\phi=-\psi\label{eq:phi-psi-constraint}\end{equation}
which reflects the conformal nature of the theory in the large $k$
limit. 

The $0i$ component of the Einstein equation simply implies that the
vector perturbations of the metric are absent in the small wavelength
limit. It follows from Eqs (\ref{eq:psi-phi}), (\ref{eq:phi-psi-constraint}),
(\ref{eq:D-E}) and (\ref{eq:consistency}) that $\chi$, $\psi$
and $\phi$ must satisfy the wave equations\[
\square\chi=\square\psi=\square\phi=0.\]

In summary, our model in the ultraviolet limit contains $N$ nearly
massless vector fields evolving according to wave equations, the free
gravitational waves propagating in a locally Minkowski space, and
scalar perturbations of the metric sourced by quantum fluctuations
of the vector fields described by Eqs (\ref{eq:psi-BdB}) and (\ref{eq:phi-psi-constraint}).
To quantize the theory in this limit, we have to quantize $3N$ independent
harmonic oscillators with an ultra-relativistic dispersion relation
$\omega_{k}\approx k$. Due to the Heisenberg uncertainty principle
the amplitude of quantum fluctuation $\delta B_{k}\approx\frac{1}{\sqrt{\omega_{k}}}\approx\frac{1}{\sqrt{k}}$.
From Eq. (\ref{eq:psi-BdB}) the amplitude of $\psi$ for randomly
oriented fluctuations is of order $\sqrt{\frac{N}{k}}B$.

\section{Superhorizon perturbations}

The evolution of perturbations on the superhorizon scales is a bit
more involved. To simplify the analysis and ensure stability of gravitational
waves \cite{GMV-grav} we assume the small field values $B\ll\frac{1}{\sqrt{N}}$,
and a large number of the fields $N\gg1$. In this limit the usual
slow-roll conditions imply (see Ref. \cite{GMV-grav} ) \begin{eqnarray*}
\frac{V_{,I}}{V}B^{2} & \ll & 1\\
\frac{V_{,II}}{V}B^{4} & \ll & 1.\end{eqnarray*}
and the terms with $V_{,II}$ are highly suppressed. 

The equations for superhorizon modes ($k\ll1$) can be obtained from
Eqs. (\ref{eq:EOM01}) and (\ref{eq:EOM02}) by neglecting the terms
involving spacial derivatives. The corresponding equations are 

\begin{equation}
\left(\phi+\psi\right)_{,i}A_{i}^{\prime}+2{\cal H}^{2}\left(\delta A_{0}+{\cal V}_{i}A_{i}\right)=0\label{eq:EOM1}\end{equation}

\begin{eqnarray}
 & \left(\psi^{\prime\prime}+2{\cal H}\psi^{\prime}-4\phi a^{2}V_{,I}\right)A_{j}=\nonumber \\
 & -\delta A_{j}^{\prime\prime}-\left(2a^{2}V_{,I}-\frac{a^{\prime\prime}}{a}\right)\delta A_{j}+\delta A_{0,j}^{\prime}+\bigtriangleup{\delta A}_{j}^{T}+\left({\cal V}_{j,k}-h_{jk}^{\prime}\right)A_{k}^{\prime}\label{eq:EOM2}\end{eqnarray}
where the background equations (\ref{eq:field}), (\ref{eq:einstein})
and (\ref{eq:einstein2}) have been used. %
\footnote{Note, that in contrast to the scalar field inflation the perturbations
of different types (scalar, vector and tensor) do not decouple. The
reason is very clear: even in the first order in perturbations one
can couple vector and tensor quantities to the background fields \cite{Armendariz}. %
}

\subsection{Decaying modes}

The temporal component of the Einstein equations to the leading order
is

\begin{equation}
\phi=\frac{V_{,I}}{V}\left({\psi B}^{2}+\frac{1}{3a^{2}}\left\langle A_{j}\left(\delta A_{j}-\frac{1}{{\cal H}}\delta A_{j}^{\prime}\right)\right\rangle \right)-\frac{\psi'}{{\cal H}}.\label{eq:phi-psi}\end{equation}
where the relation $A_{i}^{\prime}-{\cal H}A_{i}=aB_{i}^{\prime}=-\frac{2}{3}a^{2}V_{,I}A_{i}$
was used. The Eqs. (\ref{eq:EOM1}) and (\ref{eq:EOM2}) can be further
simplified

\begin{eqnarray}
{\delta A}_{0} & = & -{\cal V}_{i}A_{i},\label{eq:constraint1}\end{eqnarray}

\begin{eqnarray}
\left(\psi^{\prime\prime}+2{\cal H}\psi^{\prime}+4a^{2}V_{,I}\frac{V_{,I}}{V}\left(B^{2}\psi+\frac{1}{3a^{2}}\left\langle A_{i}\left(\delta A_{i}-\frac{1}{{\cal H}}\delta A_{i}^{\prime}\right)\right\rangle \right)\right)A_{j}\nonumber \\
+\delta{A''}_{j}+\left(2a^{2}V_{,I}-\frac{a''}{a}\right)\delta A_{j}+{h'}_{jk}{A'}_{k} & = & 0\label{eq:motion_perturbed}\end{eqnarray}
and the divergence of Eq. (\ref{eq:motion_perturbed}) implies\begin{eqnarray}
\left(\psi_{,j}^{\prime\prime}+2{\cal H}\psi_{,j}^{\prime}+4a^{2}V_{,I}\frac{V_{,I}}{V}\left(B^{2}\psi_{,j}+\frac{1}{3a^{2}}\left\langle A_{i}\left(\delta A_{i}-\frac{1}{{\cal H}}\delta A_{i}^{\prime}\right)_{,j}\right\rangle \right)\right)A_{j}\nonumber \\
+\bigtriangleup\chi''+\left(2a^{2}V_{,I}-\frac{a''}{a}\right)\bigtriangleup\chi & = & 0.\label{eq:motion_perturbed1}\end{eqnarray}

From the consistency condition ($\nabla_{\mu}\delta A^{\mu}=0$):
\begin{eqnarray}
\bigtriangleup\chi & = & {\delta A}_{0}^{\prime}+2{\cal H}\delta A_{0}=0\label{eq:constraint2}\end{eqnarray}
and \begin{eqnarray}
\psi^{\prime\prime}+2{\cal H}\psi^{\prime}+4a^{2}V_{,I}\frac{V_{,I}}{V}\left(B^{2}\psi+\frac{1}{3a^{2}}\left\langle A_{i}\left(\delta A_{i}-\frac{1}{{\cal H}}\delta A_{i}^{\prime}\right)\right\rangle \right) & = & 0.\label{eq:motion_perturbed2}\end{eqnarray}
From Eqs. (\ref{eq:constraint1}) and (\ref{eq:constraint2}) we conclude
that in the large wavelength limit the temporal $\delta A_{0}$ and
longitudinal $\chi$ components of vector fields as well as vector
perturbations of the metric ${\cal V}_{i}$ are exponentially suppressed.

\subsection{Tensor-vector mixing}

In the remainder of the section we consider the evolution of only
transverse components of the vector fields $\delta A=\delta A_{i}^{T}$
weakly coupled to the scalar and tensor perturbations of the metric,
such that the coupling can be considered perturbativly. 

The vectorial part of Eq. (\ref{eq:motion_perturbed}) 

\begin{equation}
\delta A_{j}^{T\prime\prime}+\left(2a^{2}V_{,I}-\frac{a^{\prime}}{a}\right)\delta A_{j}^{T}+h_{jk}^{\prime}A_{k}^{\prime}=0\label{eq:field_vector_perturbations}\end{equation}
and the spatial part of the Einstein equations at the leading order 

\begin{eqnarray}
\frac{a^{2}}{2N}\left(h_{ik}^{\prime\prime}+2{\cal H}h_{ik}^{\prime}\right) & = & \left\langle -\frac{1}{3}\left(\psi^{\prime\prime}B^{2}+2{\cal H}B^{2}\psi^{\prime}+\left(A_{j}\delta A_{j}\right)''\right)\delta_{k}^{i}+2{A'}_{i}\left(2{\cal H}\delta A_{k}+\delta{A'}_{k}\right)\right\rangle ,\label{eq:scalar-vector-tensor-mixture}\end{eqnarray}
can be reduced to

\begin{eqnarray}
a^{2}h_{ik}^{\prime} & = & 2N\left\langle \left(A_{i}\delta A_{k}-\frac{1}{3}A_{j}\delta A_{j}\delta_{k}^{i}\right)^{\prime}\right\rangle ,\label{eq:tensor-vector}\end{eqnarray}
where the constant of integration is suppressed by the scale factor
and can be ignored. 

It is convenient to rewrite the two relevant equations (\ref{eq:field_vector_perturbations})
and (\ref{eq:tensor-vector}) in terms of $B$ fields and in physical
time coordinates

\begin{eqnarray}
\frac{d}{dt}\left\langle B_{i}\delta B_{k}-\frac{1}{3}B_{j}\delta B_{j}\delta_{k}^{i}\right\rangle +2H\left\langle B_{i}\delta B_{k}-\frac{1}{3}B_{j}\delta B_{j}\delta_{k}^{i}\right\rangle  & = & \frac{1}{2N}\dot{h_{ik}}\label{eq:TV1}\\
\ddot{\delta B_{i}^{T}}+3H\dot{\delta B_{i}^{T}}+2V_{,I}\delta B_{i}^{T} & = & -\dot{h_{ik}}\left(\dot{B_{k}}+HB_{k}\right).\label{eq:TV2}\end{eqnarray}
In the limit of small mixing between vector fields and gravitational
waves the solution of Eq. (\ref{eq:TV1}) reads as \begin{equation}
h_{ik}=2N\left\langle B_{i}\delta B_{k}-\frac{1}{3}B_{j}\delta B_{j}\delta_{k}^{i}\right\rangle +2NH\int_{0}^{t}\left\langle B_{i}\delta B_{k}-\frac{1}{3}B_{j}\delta B_{j}\delta_{k}^{i}\right\rangle dt+C_{ik}.\label{eq:h_ik}\end{equation}

If the constant of integration $C_{ik}$ (which is determined by the
initial conditions ) dominates, the evolution of gravitation waves
resembles the usual freeze-out of the super horizon modes. According
to the Eq. (\ref{eq:TV2}) in the small coupling limit, the transverse
modes $\delta B^{T}$ change very slowly, but the behavior might change
in the long run when the second term in Eq. (\ref{eq:h_ik}), which
scales as $\sqrt{t}$, begins to dominate. Thus, the spectrum of gravitational
waves on the very large scales depends crucially on the initial conditions
at the horizon crossing. Evaluation of the horizon size modes is a
challenging task which is beyond the scope of current discussion.

\subsection{Scalar-vector mixing}

In analogy to the previous subsection we rewrite Eqs. (\ref{eq:psi-phi})
and (\ref{eq:motion_perturbed2}) in terms of $B$ fields in physical
time coordinates \begin{eqnarray}
\ddot{\psi}+3H\dot{\psi}+4V_{,I}\left(\frac{V_{,I}}{V}B^{2}\right)\left(\psi-\frac{\left\langle B_{j}\dot{\delta B_{j}}\right\rangle }{3B^{2}}\right) & = & 0\label{eq:SV1}\\
\ddot{\psi}+\left(3H+4\frac{V_{,I}}{H}\right)\dot{\psi}+4V_{,I}\phi & = & 0.\label{eq:SV2}\end{eqnarray}
In the slow-roll regime the scalar perturbation $\psi$ varies slowly
as a function of physical time. By neglecting the time-dependence
of the coefficients we can approximate the solution of Eq. (\ref{eq:SV1})
as\begin{equation}
\psi=\psi_{0}\, e^{-\frac{4V_{,I}}{3H}\left(\frac{V_{,I}}{V}B^{2}\right)t}+\frac{2V_{,I}}{9HB^{2}}\left\langle B_{j}\delta B_{j}\right\rangle ,\label{eq:psi}\end{equation}
where $\psi_{0}$ is the initial amplitude of scalar perturbations
at the horizon crossing and the second term is obtained from Eq. (\ref{eq:TV2})
in the limit of weak coupling of gravitational waves. The solution
of Eq. (\ref{eq:SV2}) is \[
\phi\approx\left(1+\frac{4V_{,I}}{3H^{2}}\right)\left(\frac{V_{,I}}{V}B^{2}\right)C\, e^{-\frac{4V_{,I}}{3H}\left(\frac{V_{,I}}{V}B^{2}\right)t}\]
which means that at very large times the scalar perturbation $\psi\approx\frac{1}{3}B^{-2}\left\langle B_{j}\dot{\delta B_{j}}\right\rangle $and
$\phi\approx0$ is exponentially suppressed. 

One should note that the decaying exponent is very small and for realistic
scenarios the first term in Eq. (\ref{eq:psi}) gives the dominant
contribution. For example, the initial spectrum of perturbations in
vector inflation with Coleman-Weinberg potential is such that $\psi_{0}\approx\frac{N}{6}\left\langle B_{i}\delta B_{i}\right\rangle \gg\frac{2V_{,I}}{9HB^{2}}\left\langle B_{j}\delta B_{j}\right\rangle $
and the solution of Eq. (\ref{eq:SV2}) is greatly simplified: \[
\phi\approx\frac{V_{,I}}{V}B^{2}\psi.\]
The slow roll parameter $\frac{V_{,I}}{V}B^{2}$ eventually becomes
of order one and the standard evolution of the decoupled scalar perturbations
proceeds with $\psi\approx\phi$, where all of the results of the
standard cosmology apply.

\section{Conclusions}

The analysis of the cosmological perturbation in models of inflation
driven by vector fields proved to be a challenging task. One issue
was recently raised in \cite{Peloso} where the stability of the longitudinal
component was questioned. The problem requires a much more detailed
examination before definite conclusions could be drawn. In the Appendix
C we only give some simple arguments which invalidate the conclusions
of Refs. \cite{Peloso,Peloso2}, and a lot more technical analysis
will be given elsewhere \cite{GMV-ghost}. 

Another issue, which is the main subject of the current paper, is
related to non-trivial coupling between scalar, vector and tensor
modes \cite{Armendariz}. It was shown that the coupling is generically
suppressed for the small fields models with isotropic backgrounds which were previously proved
to be stable to tensor perturbations \cite{GMV-grav}. Nevertheless,
the mixing of the different modes could still lead to detectable correlations
of the scalar and tensor modes in the CMB. The evolution of the superhorizon
scalar modes is weakly influenced by the trace of $\left\langle B_{i}\delta B_{k}\right\rangle $,
whose traceless part sources the tensor perturbations of the metric.
At the same time, the temporal and longitudinal components of vector
fields as well as vector perturbations of the metric remain suppressed
throughout the evolution. 

We conclude that the models of vector inflation are viable cosmological
scenarios which are compatible with all of the experimental tests
performed so far. On the other hand, the inflation driven by vector
fields could give rise to very distinct observational signatures,
such as anisotropic expansion, to be tested by future experiments.
(Although our current analysis was restricted to isotropic backgrounds.)
In addition, a possible detection of the correlations between scalar
and tensor modes of the CMB could indicate the vectorial nature of
the inflationary stage. 

Generally, new inflationary models are expected to give
a too small tensor-to-scalar ratio. The problem can be solved with a nontrivial kinetic
term \cite{VikMukh}, but in the case of vector inflation we have an alternative solution because the
tensor perturbation can hopefully be enlarged due to the coupling of
modes. Of course, some more detailed analysis is needed for definite conclusions
to be drawn concerning the observational possibilities. However, one can also hope
to facilitate the tensor mode detection in CMB due to potential correlation with the easily
observed scalar perturbations.

\begin{acknowledgments}
The authors are grateful to Viatcheslav Mukhanov, Misao Sasaki and
Alexander Vikman for useful discussions. This work was supported in
part by TRR 33 {}``The Dark Universe'' and the Cluster of Excellence
EXC 153 {}``Origin and Structure of the Universe''. 
\end{acknowledgments}

\appendix

\section{Energy-momentum tensor}

The energy momentum tensor for vector inflation was first derived
in Ref. \cite{GMV} \begin{multline}
T_{\beta}^{\alpha}=\frac{1}{4}F^{\mu\nu}F_{\mu\nu}\delta_{\beta}^{\alpha}-F^{\alpha\gamma}F_{\beta\gamma}+\left(2\frac{dV}{dI}+\frac{R}{6}\right)A^{\alpha}A_{\beta}+V\delta_{\beta}^{\alpha}\label{eq:emt1}\\
+\frac{1}{6}\left(R_{\beta}^{\alpha}-\frac{1}{2}\delta_{\beta}^{\alpha}R\right)A^{\gamma}A_{\gamma}+\frac{1}{6}\left(\delta_{\beta}^{\alpha}\square-\bigtriangledown^{\alpha}\bigtriangledown_{\beta}\right)A^{\gamma}A_{\gamma}.\end{multline}
Most of the terms on the right hand side remain important throughout
the evolution and thus, it is instructive to obtain exact first order
expressions for each of these terms separately %
\footnote{In the linearized theory the Einstein tensor is such that ($G_{i}^{k}=G_{k}^{i}$),
since $G_{i}^{k}=g^{k\mu}G_{\mu i}$ and $G_{\mu\nu}^{(0)}$ is diagonal.
On the other hand, $T_{\mu\nu}^{(n)}$ is not generally symmetric
and ($T_{i}^{k(n)}\neq T_{k}^{i(n)}$) due to the non-symmetric terms
(i.e. $h_{ij}A_{j}^{(n)}A_{k}^{(n)}$), but the overall sum $\sum_{n}T_{i}^{k(n)}$
over isotropic configurations of the fields has to be symmetric.%
}\[
\frac{1}{4}F^{\mu\nu}F_{\mu\nu}\delta_{\beta}^{\alpha}=-\frac{1}{2}a^{-4}\left({A'}_{i}^{2}\left(1+2\left(\psi-\phi\right)\right)+2{A'}_{i}{\delta A}_{i}^{\prime}-2{A'}_{i}{\delta A}_{0,i}+h_{ij}{{A'}_{i}A'}_{j}\right)\delta_{\beta}^{\alpha}\]

\[
-F^{0\gamma}F_{0\gamma}=a^{-4}\left({A'}_{i}^{2}\left(1+2\left(\psi-\phi\right)\right)+2{A'}_{i}{\delta A}_{i}^{\prime}-2{A'}_{i}{\delta A}_{0,i}+h_{ij}{{A'}_{i}A'}_{j}\right)\]

\[
-F^{0\gamma}F_{k\gamma}=a^{-4}{A'}_{i}{\delta A}_{[i,k]}^{T}\]

\[
-F^{i\gamma}F_{k\gamma}=a^{-4}\left({{A'}_{i}A'}_{k}+2{{A'}_{i}A'}_{k}\left(\psi-\phi\right)+{A'}_{i}{\delta A}_{k}^{\prime}-{A'}_{i}{\delta A}_{0,k}+{A'}_{k}{\delta A}_{i}^{\prime}-{A'}_{k}{\delta A}_{0,i}+{A'}_{k}{A'}_{j}h_{ij}\right)\]

\[
V\delta_{\beta}^{\alpha}=\left(V+\frac{1}{a^{2}}V_{,I}\left({2\psi A}_{i}^{2}+A_{i}\left(2{\delta A}_{i}^{}+h_{ij}A_{j}\right)\right)\right)\delta_{\beta}^{\alpha}\]

\[
\frac{1}{6}G_{0}^{0}A^{\gamma}A_{\gamma}=\frac{1}{6}a^{-4}\left({\cal H}^{2}\left(-3A_{i}^{2}\left(1+2\left(\psi-\phi\right)\right)-3A_{i}\left(h_{ij}A_{j}+2{\delta A}_{i}^{}\right)\right)+A_{i}^{2}\left(6{\cal H}\psi'-2\bigtriangleup\psi\right)\right)\]

\[
\frac{1}{6}G_{k}^{0}A^{\gamma}A_{\gamma}=\frac{1}{6}a^{-4}A_{i}^{2}\left(-\frac{1}{2}\bigtriangleup{\cal V}_{k}-{2\psi'}_{,k}-2{\cal H}\phi_{,k}\right)\]

\begin{eqnarray*}
\frac{1}{6}G_{k}^{i}A^{\gamma}A_{\gamma} & = & \frac{1}{6}a^{-4}A_{j}^{2}\left(\left[\left({\cal H}^{2}-2\frac{a''}{a}\right)\left(1+2\left(\psi-\phi\right)\right)+2{\cal H}\left(2\psi'+\phi'\right)+\bigtriangleup\left(\phi-\psi\right)+2\psi''\right.\right.\\
 &  & +\left.\left({\cal H}^{2}-2\frac{a''}{a}\right)\left(2{\delta A}_{j}+A_{l}h_{jl}\right)\right]\delta_{k}^{i}\\
 &  & +\left.\left(\psi-\phi\right)_{,ik}+{\cal H}\left({\cal V}_{\{ i,k\}}-{h'}_{ik}\right)+\frac{1}{2}{{\cal V}'}_{\{ i,k\}}-\frac{1}{2}{h''}_{ik}+\frac{1}{2}\bigtriangleup h_{ik}\right)\end{eqnarray*}

\[
\left(2V_{,I}+\frac{R}{6}\right)A^{0}A_{0}=0\]

\[
\left(2V_{,I}+\frac{R}{6}\right)A^{0}A_{k}=a^{-4}\left(2a^{2}V_{,I}-\frac{a''}{a}\right)A_{k}\left({\delta A}_{0}+A_{i}{\cal V}_{i}\right)\]

\begin{eqnarray*}
\left(2V_{,I}+\frac{R}{6}\right)A^{i}A_{k} & = & a^{-4}\left(\left(\frac{a''}{a}-2a^{2}V_{,I}\right)\left(A_{i}A_{k}+A_{\{ i}{{\delta A}_{k}}_{\}}+h_{ij}A_{k}A_{j}\right)\right.\\
 &  & +A_{i}A_{k}\left(-4a^{2}\psi V_{,I}-2V_{,II}\left({2\psi A}_{l}^{2}+A_{l}\left(2{\delta A}_{l}+h_{lj}A_{j}\right)\right)\right.\\
 &  & +\left.\left.\frac{1}{3}\bigtriangleup\left(2\psi-\phi\right)-{\cal H}\left(3\psi'+\phi'\right)+2\frac{a''}{a}\left(\psi-\phi\right)-\psi''\right)\right)\end{eqnarray*}

\begin{eqnarray*}
\frac{1}{6} & \left(\square\delta_{0}^{0}-\bigtriangledown^{0}\bigtriangledown_{0}\right)A^{\gamma}A_{\gamma}= & a^{-4}\left({\cal H}A_{i}\left({\cal H}A_{i}-{A'}_{i}\right)\left(1+2\left(\psi-\phi\right)\right)-2{\cal H}\psi'A_{i}^{2}+\psi'A_{i}{A'}_{i}+\frac{1}{3}A_{i}^{2}\bigtriangleup\psi\right.\\
 &  & +2{\cal H}^{2}A_{i}{\delta A}_{i}-{\cal H}\left(A_{i}{\delta A}_{i}\right)^{\prime}+\frac{1}{3}A_{i}\bigtriangleup{\delta A}_{i}\\
 &  & +\left.{\cal H}^{2}h_{ij}A_{i}A_{j}-\frac{1}{2}{\cal H}\left(h_{ij}A_{i}A_{j}\right)^{\prime}+\frac{1}{6}A_{i}A_{j}\bigtriangleup h_{ij}\right)\end{eqnarray*}

\begin{eqnarray*}
-\frac{1}{6}\bigtriangledown^{0}\bigtriangledown_{k}A^{\gamma}A_{\gamma} & = & \frac{1}{6}a^{-4}\left(2{\cal H}A_{i}^{2}\left(\phi-3\psi\right)+2A_{i}^{2}\psi'+2A_{i}{A'}_{i}\left(2\psi-\phi\right)+2\left(A_{i}{\delta A}_{i}\right)^{\prime}-6{\cal H}A_{i}{\delta A}_{i}\right.\\
 &  & +\left.\left(h_{ij}A_{i}A_{j}\right)^{\prime}-3{\cal H}h_{ij}A_{i}A_{j}\right)_{,k}\end{eqnarray*}

\begin{eqnarray*}
\frac{1}{6}\left(\square\delta_{k}^{i}-\bigtriangledown^{i}\bigtriangledown_{k}\right)A^{\gamma}A_{\gamma} & = & \frac{1}{6}a^{-4}\left(\left[2\left(\frac{a''}{a}-2{\cal H}^{2}\right)A_{j}^{2}-2{A'}_{j}^{2}-2A_{j}{A''}_{j}+6{\cal H}A_{j}{A'}_{j}\left(1+2\left(\psi-\phi\right)\right)\right.\right.\\
 &  & +2{\cal H}A_{j}^{2}\left(\psi'-\phi'\right)+2A_{j}{A'}_{j}\left(\phi'-2\psi'\right)-2A_{j}^{2}\psi''+2A_{j}^{2}\bigtriangleup\psi\\
 &  & -2\left(A_{j}{\delta A}_{j}\right)^{\prime\prime}+6{\cal H}\left(A_{j}{\delta A}_{j}\right)^{\prime}+4\left(\frac{a''}{a}-2{\cal H}^{2}\right)A_{j}{\delta A}_{j}+2A_{j}\bigtriangleup{\delta A}_{j}\\
 &  & \left.-\left(h_{jl}A_{j}A_{l}\right)^{\prime\prime}+3{\cal H}\left(h_{jl}A_{j}A_{l}\right)^{\prime}+2\left(\frac{a''}{a}-2{\cal H}^{2}\right)h_{jl}A_{j}A_{l}+A_{j}A_{l}\bigtriangleup h_{jl}\right]\delta_{k}^{i}\\
 &  & \left.-\left(2A_{j}^{2}\psi+2A_{j}{\delta A}_{j}+h_{jl}A_{j}A_{l}\right)_{,ik}+\left(A_{j}{A'}_{j}-{\cal H}A_{j}^{2}\right)\left({\cal V}_{\{ i,k\}}-{h'}_{ik}\right)\right)\end{eqnarray*}
By combining all of these terms together we obtain \begin{eqnarray*}
a^{4}T_{0}^{0} & = & \left(\frac{1}{2}{A'}_{i}^{2}+\frac{1}{2}{\cal H}^{2}A_{i}^{2}-{\cal H}A_{i}{A'}_{i}\right)\left(1+2\left(\psi-\phi\right)\right)-{\cal H}\psi'A_{i}^{2}+\psi'A_{i}{A'}_{i}-{A'}_{i}{\delta A}_{0,i}\\
 &  & +{\cal H}^{2}A_{i}{\delta A}_{i}+a^{4}V+{2a}^{2}V_{,I}\left({\psi A}_{i}^{2}+A_{i}{\delta A}_{i}+\frac{1}{2}h_{ij}A_{i}A_{j}\right)-{\cal H}\left(A_{i}{\delta A}_{i}\right)^{\prime}\\
 &  & +{A'}_{i}{{\delta A}_{i}}^{\prime}+\frac{1}{3}A_{i}\bigtriangleup{\delta A}_{i}+\frac{1}{2}{\cal H}^{2}h_{ij}A_{i}A_{j}-\frac{1}{2}{\cal H}\left(h_{ij}A_{i}A_{j}\right)^{\prime}+\frac{1}{2}h_{ij}{{A'}_{i}A'}_{j}+\frac{1}{6}A_{i}A_{j}\bigtriangleup h_{ij}\end{eqnarray*}

\begin{eqnarray*}
a^{4}T_{k}^{0} & = & {A'}_{i}{\delta A}_{[i,k]}^{T}-\left(\frac{a''}{a}-2a^{2}V_{,I}\right)A_{k}\left({\delta A}_{0}+A_{i}{\cal V}_{i}\right)-\frac{1}{12}A_{i}^{2}\bigtriangleup{\cal V}_{k}\\
 &  & +\left(-{\cal H}A_{i}^{2}\psi+\frac{1}{3}A_{i}{A'}_{i}\left(2\psi-\phi\right)+\frac{1}{3}\left(A_{i}{\delta A}_{i}\right)^{\prime}-{\cal H}A_{i}{\delta A}_{i}+\frac{1}{6}\left(h_{ij}A_{i}A_{j}\right)^{\prime}-\frac{1}{2}{\cal H}h_{ij}A_{i}A_{j}\right)_{,k}\end{eqnarray*}

\begin{eqnarray*}
a^{4}T_{k}^{i} & = & \left[\left(-\frac{5}{6}{A'}_{j}^{2}-\frac{1}{2}{\cal H}^{2}A_{j}^{2}-\frac{1}{3}A_{j}{A''}_{j}+{\cal H}A_{j}{A'}_{j}\right)\left(1+2\left(\psi-\phi\right)\right)\right.\\
 &  & +\frac{1}{6}A_{j}^{2}\bigtriangleup\left(\phi+\psi\right)+{\cal H}A_{j}^{2}\psi'+\frac{1}{3}A_{j}{A'}_{j}\left(\phi'-2\psi'\right)+{A'}_{j}{\delta A}_{0,j}\\
 &  & +\left({\cal H}A-{A'}_{j}\right){{\delta A}_{j}}^{\prime}-\frac{1}{3}\left(A_{j}{\delta A}_{j}\right)^{\prime\prime}+{\cal H}\left(A'{}_{j}-{\cal {\cal H}}A_{j}\right){\delta A}_{j}+\frac{1}{3}A_{j}\bigtriangleup{\delta A}_{j}\\
 &  & -\frac{1}{6}\left(h_{jl}A_{j}A_{l}\right)^{\prime\prime}+\frac{1}{2}{\cal H}\left(h_{jl}A_{j}A_{l}\right)^{\prime}-\frac{1}{2}{\cal H}^{2}h_{jl}A_{j}A_{l}+\frac{1}{6}A_{j}A_{l}\bigtriangleup h_{jl}-\frac{1}{2}h_{jl}{{A'}_{j}A'}_{l}+a^{2}V_{,I}h_{jl}A_{j}A_{l}\\
 &  & +\left.a^{4}V+2a^{2}V_{,I}\left({\psi A}_{l}^{2}+A_{l}{\delta A}_{l}\right)\right]\delta_{k}^{i}-2A_{i}A_{k}V_{,II}\left({2\psi A}_{l}^{2}+A_{l}\left(2{\delta A}_{l}+h_{lj}A_{j}\right)\right)\\
 &  & +A_{i}A_{k}\left(\frac{1}{3}\bigtriangleup\left(2\psi-\phi\right)-{\cal H}\left(3\psi'+\phi'\right)-2\frac{a''}{a}\phi-\psi''\right)\\
 &  & +\left(\frac{a''}{a}-2a^{2}V_{,I}\right)A_{i}A_{k}\left(1+2\psi\right)+{A'}_{i}A_{k}^{\prime}\left(1+2\left(\psi-\phi\right)\right)\\
 &  & -\frac{1}{6}A_{j}^{2}\left(\psi+\phi\right)_{,ik}+\left(\frac{a''}{a}-2a^{2}V_{,I}\right)\left(A_{\{ i}{\delta A}_{k\}}+h_{ij}A_{k}A_{j}\right)\\
 &  & +{A'}_{i}{\delta A}_{k}^{\prime}+{A'}_{k}\delta A_{i}^{\prime}-\left(A'_{i}{\delta A}_{0,k}+{A'}_{k}{\delta A}_{0,i}\right)-\left(\frac{1}{3}A_{j}{\delta A}_{j}+\frac{1}{6}h_{jl}A_{j}A_{l}\right)_{,ik}\\
 &  & +\frac{1}{6}A_{j}{A'}_{j}\left({\cal V}_{\{ i,k\}}-{h'}_{ik}\right)+\frac{1}{12}A_{j}^{2}\left({{\cal V}'}_{\{ i,k\}}-{h''}_{ik}\right)+\frac{1}{12}A_{j}^{2}\bigtriangleup h_{ik}+{A'}_{k}{A'}_{j}h_{ij}.\end{eqnarray*}

\section{Adiabatic perturbations}

Consider the scalar metric perturbations accompanied by adiabatic
modes $\delta A_{i}=fA_{i}$. It follows from equations of motion
that \begin{equation}
\bigtriangledown_{\nu}\left(2V_{,I}+\frac{R}{6}\right)A^{\nu}=0\label{eq:consistency}\end{equation}
 which is simplified in the slow roll limit to \[
\delta A_{0}'+2{\cal H}\delta A_{0}\approx\delta A_{j,j}=f_{,j}A_{j}.\]
This means that we should also allow for $\delta A_{0}\neq0$ perturbations
for the consistency relation of Eq. (\ref{eq:consistency}) to hold.

>From Eq. (\ref{eq:EOM02}) we obtain

\begin{eqnarray}
-2\phi{A''}_{i}+\left(f'-\phi'-\psi'\right){A'}_{i}-\bigtriangleup fA_{i}+f_{,ij}A_{j}-{\delta A_{0}}_{,i}'\label{eq:EOM_adiabatic}\\
+4V_{,II}A_{j}^{2}\left(\psi+f\right)A_{i}+\left(\psi''+\frac{1}{3}\bigtriangleup\left(\phi-2\psi\right)+{\cal H}\left(\phi'+3\psi'\right)+2\phi\frac{a''}{a}\right)A_{i} & = & 0.\nonumber \end{eqnarray}
The only terms which are not collinear to $A_{i}$ are represented
by $f_{,ij}A_{j}$ (it originates from $\bigtriangleup{\delta A}_{i}^{T}=\left({\delta A}_{[i,j]}^{T}\right)_{,j}=\left(f_{,j}A_{i}-f_{,i}A_{j}\right)_{,j}$)
and ${\delta A_{0}^{\prime}}_{,i}$. Since it is not possible for
$f_{,ij}A_{j}^{(n)}-\delta A_{0,i}^{(n)\prime}\approx2{\cal H}\delta A_{0,i}^{(n)}$
to be collinear with $A_{i}^{(n)}$ for all $N$ vector fields simultaneously
without creating large anisotropies of the background, we conclude
that adiabatic perturbations can't evolve independently throughout
the evolution and must generate rotations of the background vectors
(One can also check that tensor perturbations can't compensate for
$f_{,ij}A_{j}$ term). %
\footnote{Even with different variations of the length $f^{(n)}\equiv\frac{\delta A_{i}^{(n)}}{A_{i}^{(n)}}$
for different vectors the Eq. (\ref{eq:EOM_adiabatic}) cannot be
satisfied by local perturbations without generating rotations.%
} Nevertheless, as we will see, the adiabatic modes play a central
role in the evolution of superhorizon (Section IV) and subhorizon
(Section III) perturbations when the scalar, vector and tensor perturbations
are only weakly coupled.

\section{Longitudinal mode}

Consistency of the full theory of vector inflation is a subject of
the ongoing debate mainly due to the tachyonic effective mass of the
vector fields. In Refs. \cite{Peloso,Peloso2}, the authors have noted
that the evolution of the longitudinal component $\lambda$, defined
as $A_{\mu}=\nabla_{\mu}\lambda+A_{\mu}^{T}$, is described by an
action with a wrong sign of the kinetic term%
\footnote{Throughout the paper a different definition of the longitudinal component
$\chi$ was used (See Eq. (\ref{eq:longitudinal})).%
}. As we will argue, such separation of the longitudinal component
is not generically appropriate in the full theory with gravity. 

First, note that the scalar curvature $R$ contains a surface term
with a second time derivative. For the term $RA_{\mu}A^{\mu}$ in
the action these second derivatives can be integrated out, but after
the transformation $A_{\mu}\longrightarrow\bigtriangledown_{\mu}\lambda$
the higher order derivatives are introduced into equations of motion
increasing the total number of degrees of freedom. As a result, the
Ostrogradsky quantization would necessarily lead to ghosts even if
the effective mass of the vector fields is positive. 

Secondly, the separation of the longitudinal component is not so innocent
simply because the {}``mode'' $A_{\mu}=\bigtriangledown_{\mu}\lambda$
alone does not solve the equations of motion. The arguments of the
Ref. \cite{Peloso,Peloso2} refer to a massive vector field in Minkowski
space-time, where the equations of motion resemble (at the classical
level) the three independent wave equations for $A_{i}$'s and one
constraint for $A_{0}$ so that the field evolutions are completely
under control. 

Possible instabilities could still arise on the quantum level, which
is a lot more delicate issue. Due to the broken gauge invariance we
have to treat our theory only as an effective theory and one should
expect the action to be modified at high energies. Moreover, when
the Hubble scale tachyonic mass comes from the non-minimal coupling
with gravity, treating the scalar curvature as a constant number is
not really justified at the small scales. 

Most importantly, the fields with lower indices $A_{i}$ must have
some kind of {}``instability'' simply because they \textit{should}
grow due to a purely coordinate effect \cite{GMV}. This is exactly
the type of behavior which was observed in \cite{Peloso2}. In fact,
the exponential growth of $A_{i}$'s determines an exponential decay
of $A_{0}$ which is precisely the behavior we have seen in the Section
IV. Note also that after changing the variables to $B_{i}$'s the
tachyonic mass disappears from the action, although bringing about
some new complications like time-dependent coefficients and Lorentz
symmetry breaking terms.

A detailed analysis of the stability issues will be provided in a
forthcoming article \cite{GMV-ghost}.
\end{document}